\documentclass[%
 reprint,
superscriptaddress,
 amsmath,amssymb,
 aps,
 prl,
]{revtex4-2}

\usepackage{graphicx}
\usepackage{dcolumn}
\usepackage{bm}

\begin{document}

\preprint{APS/123-QED}

\title{Mesoscopic light transport in nonlinear disordered media}

\author{Alfonso Nardi}
\author{Andrea Morandi}
\affiliation{ETH Zurich, Department of Physics, Institute for Quantum Electronics, Optical Nanomaterial Group, 8093 Zurich, Switzerland}%
\author{Romain Pierrat}
\author{\mbox{Arthur Goetschy}}
\affiliation{ESPCI Paris, PSL University, CNRS, Institut Langevin, 75005 Paris, France}%
\author{Xuanchen Li}
\affiliation{ETH Zurich, Department of Physics, Institute for Quantum Electronics, Optical Nanomaterial Group, 8093 Zurich, Switzerland}%
\author{Frank Scheffold}
\affiliation{Department of Physics, University of Fribourg, 1700 Fribourg, Switzerland}%
\author{Rachel Grange}
\affiliation{ETH Zurich, Department of Physics, Institute for Quantum Electronics, Optical Nanomaterial Group, 8093 Zurich, Switzerland}%

\date{\today}

\begin{abstract}

Nonlinear disordered media uniquely combine multiple scattering and second-harmonic generation.
Here, we investigate the statistical properties of the nonlinear light generated within such media.
We report super-Rayleigh statistics of the second-harmonic speckle intensity, and demonstrate that it is caused by the mesoscopic correlations arising in extreme scattering conditions.
The measured conductance is the lowest ever observed in an isotropically scattering 3D medium, with applications in broadband second-harmonic generation, wavefront shaping in nonlinear disordered media, and photonic computing.

\end{abstract}

\maketitle

When a coherent wave interacts with a disordered medium, it generates a complex interference phenomenon that results in the formation of speckle patterns.
These speckles, which are randomly distributed diffraction-limited grains, are ubiquitous across waves of various origins, including electromagnetic~\cite{goodman_speckle_2020}, acoustic~\cite{wagner_statistics_1983}, and matter waves~\cite{dall_observation_2011}.
Despite their different physical origins, speckles exhibit universal statistical properties that are referred to as Rayleigh statistics. 
This universality is a consequence of the very general conditions under which Rayleigh statistics emerge.
Specifically, the only requirement is that the field arises from the interference of a large number of uncorrelated waves whose phases are uniformly distributed over a $2\pi$ range~\cite{goodman_speckle_2020, carminati_principles_2021}.
Non-Rayleigh statistics can be achieved by using a spatial light modulator to induce correlations between partial waves and redistribute the intensity among the speckle pattern's grains, while preserving ergodicity~\cite{bromberg_generating_2014, liu_generation_2021, han_tailoring_2023}. 

Conversely, observing deviations from Rayleigh statistics in multiple-scattering media under conventional laser illumination is significantly more challenging.
Deviations occur only when strong scattering introduces mesoscopic correlations between partial waves, breaking ergodicity~\cite{PhysRevLett.61.834, Garcia1989, PhysRevLett.64.2787, PhysRevLett.81.5800, PhysRevLett.97.103901}. 
These correlations are associated with the existence of a finite number $g$ of open transmission channels in disordered media~\cite{dorokhov_coexistence_1984, imry_active_1986}. 
When $g$ (also known as \textit{dimensionless conductance}) is moderate, the central limit theorem does not apply, causing the total intensity in the speckle pattern to fluctuate significantly from one disorder configuration to another. 
Consequently, the intensity in each speckle grain exhibits super-Rayleigh statistics~\cite{Garcia1989, nieuwenhuizen_intensity_1995}. 
This peculiar regime has been reported in three-dimensional systems only in nanowire mats~\cite{strudley_mesoscopic_2013} and, more weakly, in isotropic ZnO scattering media~\cite{strudley_observation_2014}.

While mesoscopic transport in linear scattering media is well understood, the same cannot be said for nonlinear disordered media, a class of materials that has garnered significant interest recently.
These materials are composed of nanodomains that exhibit a second-order susceptibility tensor. 
The nonlinear characteristics enable each nanodomain to exhibit electro-optic effects and generate second-harmonic (SH) waves when illuminated with a fundamental beam~\cite{timpu_lithium_2019}.
The interference of the waves generated by the nanodomains results in efficient emission of nonlinear light without stringent conditions on the polarization and wavelength of the fundamental light, contrary to bulk crystals~\cite{boyd_nonlinear_2008}.
This phenomenon has been extensively studied in the framework of random quasi-phase-matching~\cite{baudrier-raybaut_random_2004, makeev_second_2003}, where the disorder is used to achieve efficient and broadband SH generation~\cite{fischer_broadband_2006, qiao_cavity-enhanced_2019, savo_broadband_2020, muller_modeling_2021, morandi_multiple_2022}.
Researchers investigated the fundamental properties of diffusion and weak localization in this class of media~\cite{agranovich_effects_1988, yoo_search_1989, faez_experimental_2009, valencia_weak_2009}, as well as the effect of scatterers displacement~\cite{samanta_intensity-dependent_2020, samanta_speckle_2022}.
Notably, nonlinear disordered media are also emerging as a prominent platform for photonics processing.
The characterization of the scattering tensor that defines their nonlinear input-output response allows the use of nonlinear disordered media for encryption and as all-optical logic gates~\cite{moon_measuring_2023}.
In addition, these media enables the implementation of large-scale nonlinear optical operators for photonic machine learning~\cite{wang_large-scale_2024}.
However, simplifying the analysis by assuming Rayleigh statistics, these applications might overlook the richer physics arising from strong scattering effects.

In this Letter, we examine the statistical properties of SH light generated within a strongly scattering, nonlinear disordered medium. 
Specifically, we measure the intensity fluctuations of both the fundamental and SH light under various illumination conditions. 
The histograms of speckle intensity and total transmission show deviations from Rayleigh statistics, particularly evident for SH light.
Since SH light can be generated throughout the medium and cannot be described by a linear transmission matrix, an open question is whether these observations can be interpreted in terms of an effective number $g$ of open channels for SH light. 
Here, we demonstrate that our results align well with mesoscopic transport theory based on a reduced $g$, and we explain it by analyzing the location of SH generation within the medium and the wavelength dependence of the transport mean free path of light.
This outcome has relevant applications in broadband SH generation, wavefront shaping~\cite{hsu_correlation-enhanced_2017, Bender2022}, and photonic computing~\cite{wang_large-scale_2024}.

\begin{figure}
    \centering
    \includegraphics[width=\columnwidth]{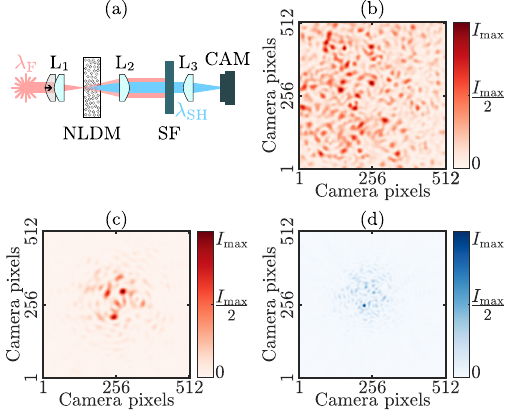}
    \caption{Experimental setup and typical speckle patterns. (a) $\lambda_\mathrm{F}=976\,\mathrm{nm}$: pulsed laser; $\lambda_\mathrm{SH}=488\,\mathrm{nm}$; $\mathrm{L}$: lens; $\mathrm{NLDM}$: nonlinear disordered medium; $\mathrm{SF}$: spectral filter; $\mathrm{CAM}$: camera.
    (b-c) Camera images of typical speckle patterns at $\lambda_F$ when the sample is illuminated with (b) out-of-focus and (c) in-focus fundamental light. 
    (d) Camera image of a typical speckle pattern from SH light generated within the medium with in-focus fundamental illumination.
    }
    \label{fig:setup}
\end{figure}

Our experimental setup is shown in Fig.~\ref{fig:setup}a.
A linearly polarized Ti:Sa pulsed laser (wavelength $\lambda_\mathrm{F} = 976\,\mathrm{nm}$, $80\,\mathrm{MHz}$ repetition rate, $100\,\mathrm{fs}$ pulse duration) is focused by an aspheric lens ($\mathrm{NA} = 0.5$) into a nonlinear disordered medium (disordered assembly of LiNbO$_3$ nanoparticles of thickness $L\approx 10\,\mathrm{\mu m}$, see Supplemental Material, Sec.~S1 for detailed information~\cite{supplementary}).
Both the lens and the disordered sample are mounted on motorized translation stages, to accurately control the position of the focal plane, as well as the position of the sample in the plane perpendicular to the beam propagation direction.
By moving the focal plane, we can control the size of the beam at the input facet of the disordered medium, a crucial parameter for the analysis of mesoscopic transport.
Changing the position of the sample allows the light traveling through the medium to interact with different realizations of disorder.
The light scattered and upconverted by the nonlinear disordered medium is collected by an objective ($\mathrm{NA} = 0.75$), and a tube lens (focal length $200\,\mathrm{mm}$) is used to image the output facet of the medium onto a scientific CMOS camera.
Finally, we use a linear polarizer to select a single polarization state, and a spectral filter (bandpass region $360-580\,\mathrm{nm}$) to remove the fundamental light when measuring the SH signal ($\lambda_\mathrm{SH} = \lambda_\mathrm{F}/2 = 488\,\mathrm{nm}$).
%
Typical speckle patterns for different illumination conditions are shown in \mbox{Fig.~\ref{fig:setup}b-d}.
Figure~\ref{fig:setup}b displays a camera image of the fundamental light when the focal plane is far from the input facet of the medium, i.e., when the illuminating beam size is large.
Figures~\ref{fig:setup}c-d report typical speckle patterns for the fundamental light (Fig.~\ref{fig:setup}c) and the SH light generated within the nonlinear disordered medium (Fig.~\ref{fig:setup}d) when the focal plane of the fundamental illumination coincides with the input facet of the sample.

In the following, we rigorously characterize the statistical properties of the transmitted light after the propagation through the nonlinear disordered medium.
In particular, we measure the intensity fluctuations, which contains strong signature of mesoscopic transport, and are not affected by absorption~\cite{chabanov_statistical_2000}.
For a given realization of disorder, we use the camera to record the intensity distribution at the output facet of the disordered sample.
From the measured speckle pattern, we obtain the transmission coefficients $T_{ab}$, which relate a speckle spot $b$ with an incoming wave $a$ (kept fixed for each statistical dataset). 
Summing over the output modes, we obtain the total transmission as $T_a = \sum_b T_{ab}$.
To collect a statistically meaningful set of measurements, we measure the intensity distributions of $10^4$ different realizations of disorder, obtained by moving the sample in the plane perpendicular to the propagation direction of the illuminating light.
From $T_{ab}$ and $T_a$, we then extract the most relevant quantities for our analysis, which are the normalized speckle intensity $s_{ab} = T_{ab} / \langle T_{ab} \rangle$ and the normalized total transmission $s_{a} = T_{a} / \langle T_{a} \rangle$, where the angle brackets denote the average over the ensemble of random configurations (see Supplemental Material, Sec.~S2 for details about the data analysis~\cite{supplementary}).

\begin{figure}
    \centering
    \includegraphics{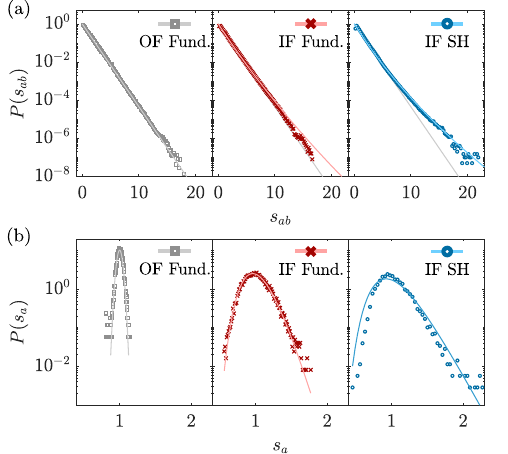}
    \caption{Intensity fluctuations statistics of fundamental light with out-of-focus (OF Fund.) and in-focus (IF Fund.) illumination, and of SH light generated within the medium with in-focus fundamental illumination (IF SH).
    (a) Histograms of normalized speckle intensity.
    Solid lines represent a negative exponential $P(s_{ab}) = -\exp(s_{ab})$ (gray), and the mesoscopic transport model for normalized speckle intensity $P_g(s_{ab})$ with conductance values $g=13$ (red) and $g=6.5$ (blue).
    (b) Histograms of normalized total transmission.
    Solid lines indicate a Gaussian fit (gray), and the mesoscopic transport model for normalized total transmission $P_g(s_{a})$, using the same conductance values as in (a), $g=13$ (red) and $g=6.5$ (blue).
    }
    \label{fig:intensity_and_TT_statistics}
\end{figure}

The results of the analysis for various illumination conditions are reported in Fig.~\ref{fig:intensity_and_TT_statistics}.
We first examine the case of fundamental light with out-of-focus illumination (Fig.~\ref{fig:intensity_and_TT_statistics}, gray).
The large size of the input beam ensures that we are in a regime of large number of open channels.
In this case, the normalized speckle intensity obeys the Rayleigh law, meaning that the probability distribution $P(s_{ab})$ is a negative exponential (gray solid line in Fig.~\ref{fig:intensity_and_TT_statistics}a).
Conversely, the probability distribution of the normalized total transmission $P(s_a)$ follows a Gaussian distribution (Fig.~\ref{fig:intensity_and_TT_statistics}b, gray).
This is a consequence of the central limit theorem, as the numerous scattering paths that are summed to obtain the total transmission are uncorrelated.

Reducing the size of the illuminating beam by aligning the focal plane with the input facet of the medium (referred to as \textit{in-focus}) increases the probability of two paths crossing during light propagation inside a disordered medium. 
When two paths cross, there is a small but non-zero probability that they will become correlated, giving rise to long-range correlations between distant speckle grains~\cite{PhysRevLett.64.2787, scheffold_observation_1997}. 
This, in turn, lowers the dimensionless conductance $g$, which quantifies the number of open channels, i.e., the number of channels that carry the majority of the light ~\cite{dorokhov_coexistence_1984, imry_active_1986, beenakker_random-matrix_1997}.
The introduced correlations change the statistical properties of the scattered light, enhancing in particular the intensity fluctuations~\cite{VanRossum99}.
Mesoscopic transport theory predicts modified probability distributions, referred to as $P_g(s_{ab})$ and $P_g(s_a)$, containing the dimensionless conductance $g$ as a single parameter (see Ref.~\cite{nieuwenhuizen_intensity_1995} and Supplemental Material, Sec.~S3~\cite{supplementary}).

The normalized speckle intensity and total transmission histograms for fundamental light with in-focus illumination is shown in red in Fig.~\ref{fig:intensity_and_TT_statistics}a and b, respectively.
Considering the speckle intensity, the slight deviation from Rayleigh distribution at large $s_{ab}$ values is consistent with previously reported observations in isotropic scattering materials~\cite{strudley_observation_2014}.
Conversely, the total transmission histogram shows a clear deviation from the Gaussian distribution predicted by the uncorrelated wave model.
Fitting the measured data with the mesoscopic transport model of Ref.~\cite{nieuwenhuizen_intensity_1995} demonstrates good agreement (Fig.~\ref{fig:intensity_and_TT_statistics}, red), resulting in a conductance value of $g=13$.

With the same in-focus position of the input lens, we measure the intensity statistics of the SH light generated within the nonlinear disordered medium.
The histogram of the normalized speckle intensity $P(s_{ab})$, shown in blue in Fig.~\ref{fig:intensity_and_TT_statistics}a, exhibits a more pronounced deviation from the Rayleigh distribution.
Consequently, fitting the data with the mesoscopic model $P_g(s_{ab})$ yields a smaller conductance ($g=6.5$).
The normalized total transmission also demonstrates increased variance (Fig.~\ref{fig:intensity_and_TT_statistics}b, in blue).
Using the same conductance value of $g=6.5$, derived from the fit of the speckle intensity histogram, we achieve a strong agreement between the theoretical model $P_g(s_a)$ and the experimental data.
This is particularly evident for large $s_a$ values, where the data points closely follows a negative exponential of the form $P(s_a) \propto e^{-g s_a}$.

\begin{figure}
    \centering
    \includegraphics{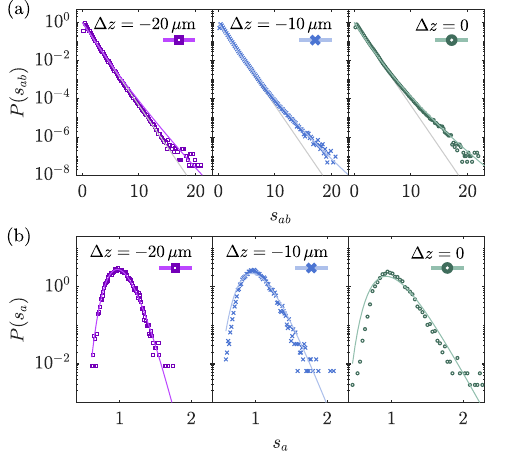}
    \caption{Intensity fluctuations statistics of SH light generated within the nonlinear disordered medium for varying distance $\Delta z$ between the focal plane of the input lens and the entry facet of the nonlinear disordered medium ($\Delta z = \{ -20\,\mathrm{\mu m},~-10\,\mathrm{\mu m},~0 \}$). 
    (a) Histograms of normalized speckle intensity.
    Solid lines represent a negative exponential $P(s_{ab}) = -\exp(s_{ab})$ (gray), and the mesoscopic transport model $P_g(s_{ab})$, with conductance values $g=16$ (violet), $g=9.5$ (blue) and $g=6.5$ (green).
    (b) Histograms of normalized total transmission. 
    Solid lines are the mesoscopic probability density functions $P_g(s_a)$ with the same conductance values as in (a), $g=16$ (violet), $g=9.5$ (blue), $g=6.5$ (green).
    }
    \label{fig:different_z_figure}
\end{figure}

To confirm the agreement with the mesoscopic transport model, we characterized the intensity fluctuations of the SH light generated within the medium at different input lens positions ($\mathrm{L_1}$ in Fig.~\ref{fig:setup}), thereby varying the fundamental spot size at the entrance of the nonlinear medium.
The results are shown in Fig.~\ref{fig:different_z_figure}.
We present histograms for three illumination conditions, based on the distance $\Delta z$ of the focal plane to the input facet of the medium: $\Delta z = -20\,\mathrm{\mu m}$, $\Delta z = -10\,\mathrm{\mu m}$, $\Delta z = 0$.
As predicted by mesoscopic transport theory~\cite{scheffold_observation_1997}, the deviation of the speckle intensity histograms (Fig.~\ref{fig:different_z_figure}a) from a negative exponential (gray solid line) becomes more evident as the focal plane of the illumination approaches the input facet of the sample.
Similarly, in the histograms of the normalized total transmission (Fig.~\ref{fig:intensity_and_TT_statistics}b), the exponential decay present at large $s_a$ becomes more pronounced for shorter $\Delta z$.
The fitted probability distributions (violet, blue and green solid lines in Fig.~\ref{fig:different_z_figure}) show excellent agreement for both the speckle intensity and the total transmission histograms, using the same conductance values for the expressions of $P_g(s_{ab})$ and $P_g(s_a)$ ($g=16$ for $\Delta z = -20\,\mathrm{\mu m}$, $g=9.5$ for $\Delta z = -10\,\mathrm{\mu m}$ and $g=6.5$ for $\Delta z = 0$).

The reduced conductance measured with the SH light is explained by the shorter transport mean free path $\ell_t$ associated to the SH wavelength ($\ell_t \approx 160\,\mathrm{nm}$ at $450\,\mathrm{nm}$, compared to $\ell_t \approx 700\,\mathrm{nm}$ at $950\,\mathrm{nm}$; see Supplemental Material, Sec.~S4 for the experimental characterization of $\ell_t$~\cite{supplementary}).
The transport mean free path is relevant for the characterization of the mesoscopic effects, because it limits the minimum conductance to a value $g_\mathrm{min}(\lambda) \approx (2\pi \ell_t/\lambda )^2$~\cite{scheffold_observation_1997, Scheffold2002}.
Considering the measured $\ell_t$, we obtain $g_\mathrm{min}(950 \, \mathrm{nm} ) \approx 11.3$ and $g_\mathrm{min}(450 \, \mathrm{nm} ) \approx 2.6$~\cite{supplementary}.
Notably, by accounting for the beam width dependence of $g$ with a finite SH illumination width of $600\,\mathrm{nm}$, consistent with our experimental parameters, we estimate a conductance of $g=6.5$ (see Supplemental Material, Sec.~S4~\cite{supplementary}).

The question remains on how the statistics can be so accurately described with mesoscopic transport theory, given that the generation of the SH light throughout the medium prevents the definition of a linear transmission matrix~\cite{moon_measuring_2023}.
The explanation lies in the fact that, due to the tight focusing, the SH generation efficiency is maximized at the entrance of the medium.
Indeed, theoretical calculations show that the SH signal is mostly generated at $z < 500\,\mathrm{nm}$ (see Fig.~\ref{fig:BBO_vs_SH_figure}a, on the left; theoretical and simulation details in Supplemental Material, Sec.~S5~\cite{supplementary}).
\begin{figure}
    \centering
    \includegraphics{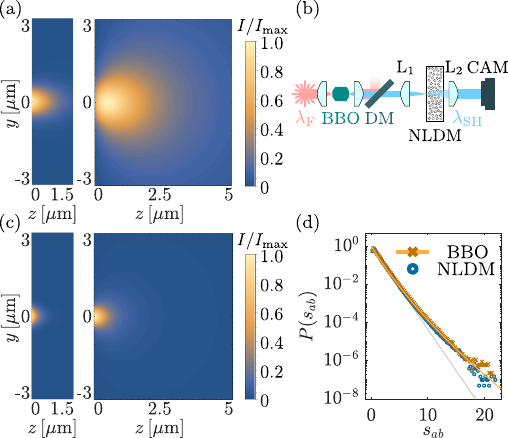}
    \caption{Comparison of SH light generated within the nonlinear disordered medium (NLDM), or externally, via a BBO crystal.
    (a, c) Calculated mean intensity profiles within the disordered medium of the sources (left) and the resulting  diffuse halos (right) for SH light generated (a) within the NLDM or (c) with a BBO crystal~\cite{supplementary}.
    (b) Experimental setup. 
    Light from a pulsed laser is upconverted with a BBO crystal ($\lambda_{F} = 976\,\mathrm{nm}$, $\lambda_\mathrm{SH} = 488\,\mathrm{nm}$).
    A dichroic mirror (DM) filters out the fundamental light, while the SH is focused by the lens $\mathrm{L_1}$ onto the NLDM.
    The scattered light is collected by an objective $\mathrm{L_2}$ and measured with a camera (CAM).
    (d) Histograms of normalized speckle intensity $P(s_{ab})$ for SH light generated with a BBO crystal (cross, yellow) or by the NLDM (circles, blue). 
    }\label{fig:BBO_vs_SH_figure}
\end{figure}
This means that, given that the nanoparticles' sizes are distributed between $100\,\mathrm{nm}$ and $400\,\mathrm{nm}$~\cite{supplementary}, the SH light is generated within the very first layers, and then undergoes linear scattering for the rest of the thickness $L\approx10\,\mathrm{\mu m}$ of the material (Fig.~\ref{fig:BBO_vs_SH_figure}a, on the right).
Therefore, under tightly focused illumination, we can assume that the system behaves similarly to a linear scattering medium.
To prove it, we compare the statistical properties of SH light generated within the nonlinear medium, and by an external $200\,\mathrm{\mu m}$ thick $\mathrm{Ba B_2 O_4}$ crystal (BBO, experimental setup depicted in Fig.~\ref{fig:BBO_vs_SH_figure}b).
Although the light generated within the nonlinear disordered medium (Fig.~\ref{fig:BBO_vs_SH_figure}a, right) penetrates slightly deeper  than the light generated by the BBO (Fig.~\ref{fig:BBO_vs_SH_figure}c, right), the sources of both concentrate at depth $z\ll L \approx 10\mu$m (Fig.~\ref{fig:BBO_vs_SH_figure}a,c, left). 
In addition, the measured histograms of $s_{ab}$ for SH light from both the nonlinear medium and the BBO (shown in Fig.~\ref{fig:BBO_vs_SH_figure}d in yellow and blue, respectively) closely match.
The similarity is expected, as the conductance is limited by the transport mean free path and the wavelength, which are identical in the two cases.
The excellent agreement between the two histograms confirms that the conductance values measured for SH light generated in the medium are determined by the scattering properties of the sample and can be accurately described by mesoscopic transport theory.

In summary, we have experimentally measured the intensity fluctuations of fundamental and SH light generated by a nonlinear disordered medium. 
We observed distinct deviations from the predictions of the uncorrelated wave model in the histograms of SH speckle intensity and total transmission. 
The measured deviations are well described by mesoscopic transport model, marking the first observation of mesoscopic transport in a nonlinear disordered medium to our knowledge.
Moreover, the measured conductance of $g=6.5$ is the lowest ever reported in an isotropically scattering 3D medium. 
We motivated the description of the nonlinear input-output response in terms of effective number $g$ of open channels by observing that the SH light is mainly generated at the input of the nonlinear disordered medium. 
Subsequently, the SH light undergoes linear scattering, allowing us to approximate the nonlinear disordered sample as a linear scattering system.
We confirmed this reasoning by comparing the speckle intensity of SH light generated either within the medium, or externally with a BBO crystal, obtaining very similar fluctuations statistics.
Our results show SH speckles with intensities more than 20 times stronger than the average, occurring with a probability two orders of magnitude higher than that of uncorrelated light. 
The increased probability of generating higher SH intensity condensed in few speckles finds useful applications in broadband SH generation from nonlinear disordered media.
Additionally, achieving a low number of open transmission modes has also significant implications for wavefront shaping, as the long-range correlations enhance the achievable control~\cite{hsu_correlation-enhanced_2017, Bender2022}.
The deviations from the uncorrelated wave model indicate that, when the fundamental light is strongly focused to achieve a higher SH signal, the wave propagation through the complex nonlinear medium cannot be described by a fully random nonlinear input-output response~\cite{moon_measuring_2023}. 
It is therefore essential to consider this effect for photonic computing applications that aim to exploit randomness to achieve large-scale nonlinear operations~\cite{wang_large-scale_2024}.

\vspace{3pt}
We acknowledge Maria Teresa Buscaglia and Vincenzo Buscaglia from the Institute of Condensed Matter Chemistry and Technologies for Energy (National Research Council, via De Marini 6, 16149 Genoa, Italy) for the synthesis of the LiNbO$_3$ nanocubes.
This research was supported by the Swiss National Science Foundation SNSF via Consolidator Grant 213713 (A.N., A.M., R.G.) and via project grant 188494 (F.S.).
R.P. and A.G. acknowledge the support of the program  Investissements d’Avenir  launched by the French Government.
A.N. and A.M. contributed equally to this work.

\bibliography{references_static}

\end{document}


\renewcommand{\theequation}{S\arabic{equation}}
\renewcommand{\thefigure}{S\arabic{figure}}
\setcounter{secnumdepth}{3}
\renewcommand\thesection{S\arabic{section}}

\title{Supplemental Material: \\ Mesoscopic light transport in nonlinear disordered media}

\author{Alfonso Nardi}\thanks{These authors contributed equally to this work.}
\author{Andrea Morandi}\thanks{These authors contributed equally to this work.}
\affiliation{ETH Zurich, Department of Physics, Institute for Quantum Electronics, Optical Nanomaterial Group, 8093 Zurich, Switzerland}%

\author{Romain Pierrat}
\author{\mbox{Arthur Goetschy}}
\affiliation{ESPCI Paris, PSL University, CNRS, Institut Langevin, 75005 Paris, France}%
\author{Xuanchen Li}
\affiliation{ETH Zurich, Department of Physics, Institute for Quantum Electronics, Optical Nanomaterial Group, 8093 Zurich, Switzerland}%
\author{Frank Scheffold}
\affiliation{Department of Physics, University of Fribourg, 1700 Fribourg, Switzerland}%
\author{Rachel Grange}
\affiliation{ETH Zurich, Department of Physics, Institute for Quantum Electronics, Optical Nanomaterial Group, 8093 Zurich, Switzerland}%

\date{\today}

\maketitle

This document provides supplementary information for the article ``Mesoscopic Light Transport in Nonlinear Disordered Media''. 
In Sec.~\ref{sec:fab}, we present the procedure for fabricating the nonlinear disordered medium. 
Section~\ref{sec:analysis} reports the detailed data analysis, and Sec.~\ref{sec:theor_expressions} presents the theoretical expressions for the probability density functions according to mesoscopic transport theory~\cite{nieuwenhuizen_intensity_1995}. 
In Sec.~\ref{sec:TMFP} we report the measurements of the sample transmittance, yielding the experimental values of the transport mean free path and the minimum conductance. 
Finally, in Sec.~\ref{sec:transport_model}, we derive  the equations that govern the propagation of the linear and SH mean intensities through the disordered slab. 

\section{Fabrication procedure}\label{sec:fab}

\begin{figure}
    \centering
    \includegraphics[width=0.5\textwidth]{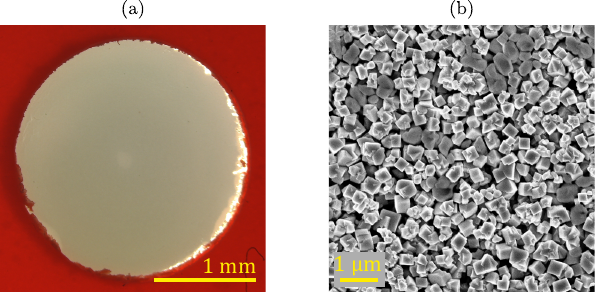}
    \caption{Nonlinear disordered sample. (a) Optical image (in reflection configuration) of the disordered sample. (b) Scanning electronic microscope image of the slab, showing a typical arrangement of the nanoparticles.}
    \label{fig:SEM_image}
\end{figure}

\begin{figure}
    \centering
    \includegraphics[width=\textwidth]{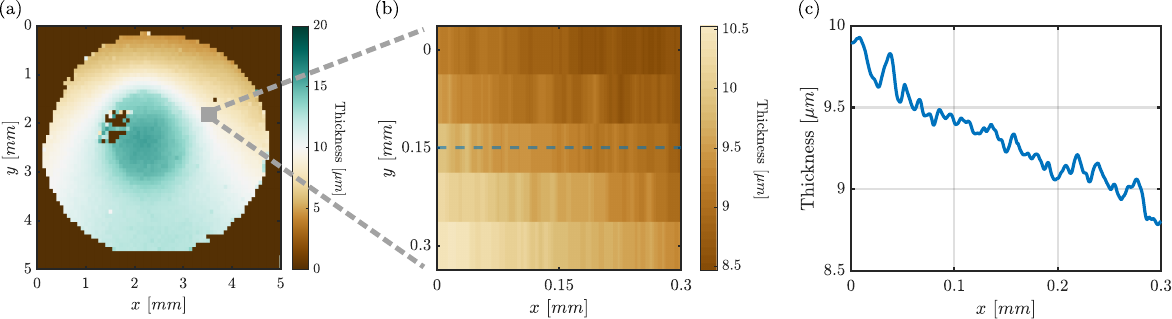}
    \caption{Thickness map of the sample, measured with a profilometer. 
    (a) Thickness map of the entire sample. 
    (b) Thickness map of the area of $300\times 300 \, \mathrm{\mu m}^2$, used for the mesoscopic transport analysis. 
    (c) Thickness profile along the blue dashed line.}
    \label{fig:thickness_map_ROI}
\end{figure}

The nonlinear disordered medium used is a slab, assembled starting from crystalline LiNbO$_3$ nanoparticles, produced by solvothermal synthesis~\cite{timpu_lithium_2019}. 
The precursor oxides, Nb2O5 (HC Starck, 99.92$\%$) and LiOH (Aldrich, 98$\%$), are dispersed in a mixture of ethylene glycol and distilled water. 
Following ultrasonication, the suspension is transferred into a polytetrafluoroethylene-coated stainless steel acid digestion bomb (model PA4748, volume $120 \, \mathrm{ml}$, Parr Instrument Company) and treated hydrothermally at $250^\circ \mathrm{C}$ for $70$ hours. 
The reaction product is then washed with water via centrifugation. 
This chemical synthesis method allows precise control over the size distribution of the nanoparticles, which range from $100$ to $400\, \mathrm{nm}$ in size, have a linear refractive index of approximately $2.3$, and exhibit negligible absorption at visible wavelengths. 
These nanoparticles feature a non-centrosymmetric hexagonal $\mathrm{R3c}$ crystal structure that enables second-harmonic (SH) generation. 
To form the slabs, we deposit the nanoparticles by drop deposition and allow solvent evaporation. 
An aqueous suspension ($1 \mathrm{wt\%}$) of the LiNbO$_3$ nanoparticles is mixed with polyvinyl alcohol in a 60:1 ratio and deposited onto a glass substrate framed with hydrophilic tape.
The sample is placed on a horizontal substrate holder and maintained at 0 °C for 24 hours. 
An optical microscope image of the resulting slab is reported in Fig.~\ref{fig:SEM_image}a.
The white opaque color is typical of multiple scattering sample that have a thickness much larger than the transport mean free path. 
The porous structure composed by randomly oriented nanoparticles and air gaps is shown in Fig.~\ref{fig:SEM_image}b, measured via 
scanning electron microscopy (SEM).
By controlling the amount of deposited nanoparticles solution, we can tune the thickness of the slab.
We measured the thickness of the sample used for the experiments in the main text by profilometry.
The resulting thickness map is shown in Fig.~\ref{fig:thickness_map_ROI}a.

\section{Data analysis}\label{sec:analysis}

\begin{figure}
    \centering
    \includegraphics[width=0.95\textwidth]{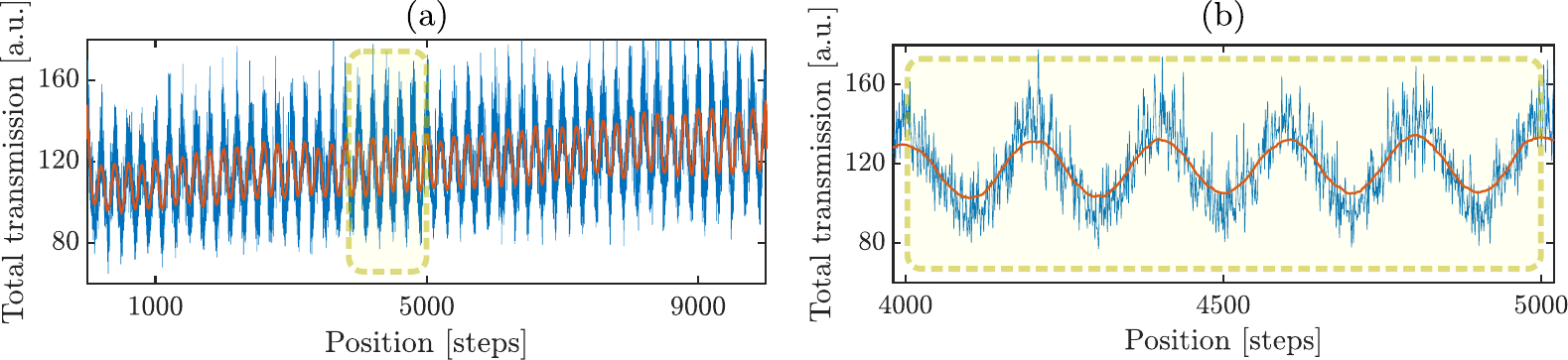}
    \caption{Total transmission fluctuations over different positions of the sample, resulting from the thicknesses gradient and fluctuations of the laser power. (a) Fluctuation over the entire dataset, and (b) close up on 1000 steps. The orange line is the result of a moving average over 100 steps.}
    \label{fig:smooth_intensity}
\end{figure}

We collected the data by measuring the speckle patterns ($\tilde{T}_{ab, i}$, where $b$ are the output modes and $a$ the fixed input modes) at $i=1 \dots N=100 \times 100$ different sample positions.
For each measurement, we moved the sample by $3 \, \mathrm{\mu m}$ using a motorized stage.
The thickness of the total scanned area of $300\times 300 \, \mathrm{\mu m}^2$, measured by profilometry, is shown in Fig.~\ref{fig:thickness_map_ROI}b.
The thickness along the blue dashed line in Fig.~\ref{fig:thickness_map_ROI}b is presented in Fig.~\ref{fig:thickness_map_ROI}c.
Due to the inhomogeneous thickness and the laser fluctuations, the total transmission ($\mathrm{TT}_i$, i.e., the sum of all the intensities on the camera pixels) exhibited periodic fluctuations, as shown in Fig.~\ref{fig:smooth_intensity}.
To account for these fluctuations, we used a moving average with a window size of 100 steps to smooth the total transmission.
This filtered total transmission $\mathrm{TT}^\mathrm{filt.}_{i}$ is shown in orange in Fig.~\ref{fig:smooth_intensity}.
We normalized each speckle pattern by the corresponding value of filtered total transmission (similar approach to Ref.~\cite{strudley_mesoscopic_2013})
\begin{equation}
    T_{ab, i} = \tilde{T}_{ab, i} \frac{ \langle \mathrm{TT}_{i} \rangle }{\mathrm{TT}^\mathrm{filt.}_{i}} \, ,
\end{equation}
where the angle brackets stands for the average over the $N$ realizations of disorder.
We then extracted the ensemble average $\langle T_{ab} \rangle$ of the speckle patterns, as
\begin{equation}
    \langle T_{ab} \rangle = \frac{1}{N} \sum_{i=1}^N T_{ab, i} \;.
\end{equation}
For the rest of the analysis, we considered only the central area (average intensity larger than $75\%$ of the maximum) of the measured images to minimize the effect of the background.
For each collected camera image, we subtracted the background and normalized the speckle patterns by the average intensity:
\begin{equation}
    s_{ab, i} = \frac{T_{ab, i}}{\langle T_{ab} \rangle} \; .
\end{equation}
We then collected all $N$ realizations into a single vector, to obtain the histogram of $s_{ab}$.
Finally, by summing all the output modes, we derived the normalized total transmission
\begin{equation}
    s_{a, i} = \sum_b s_{ab, i}
\end{equation}
and collected the $N$ realizations into the same vector $s_a$.

\section{Theoretical expressions of probability density functions}\label{sec:theor_expressions}

For clarity, we present the expressions for $P_g(s_a)$ and $P_g(s_{ab})$ used to fit the histograms, as derived in Ref.~\cite{nieuwenhuizen_intensity_1995}.
The probability density function $P_g(s_a)$ for the normalized total transmission is
\begin{equation}
    P_g(s_a) = \int_{-i\infty}^{i\infty} \frac{dx}{2\pi i} \exp \left[ x s_a - \Psi(x) \right].
\end{equation}
For broad Gaussian illumination (waist $w$ larger than than the sample thickness $L$), it was proved that the function $\Psi(x)$ takes the form
\begin{equation}
    \Psi(x) = g \int_0^1 \frac{dy}{y} \left[ \log \left( \sqrt{1 + \frac{xy}{g}} + \sqrt{\frac{xy}{g}} \right) \right]^2,
    \label{eq:Psi}
\end{equation}
where $g=k^2w^2\ell_t/3L$. We note that this definition of $g$, specific for Gaussian beam, leads to the following relation between the second moment of the normalized total transmission and the conductance $g$:
\begin{equation}
    \langle s_a^2 \rangle = 1+ \frac{1}{3g} \;.
    \label{eq:variance_and_g}
\end{equation} 
For a Gaussian beam with a waist significantly smaller than the slab thickness, an analytical expression for $P_g(s_a)$ is not available. However, as argued in Ref.~\cite{VanRossum99}, the first cumulants are expected to remain nearly unchanged, provided that the parameter $g$ is properly computed for this specific geometry. This is why we use Eq.~\eqref{eq:Psi} to fit our experimental data, with $g$ as the only fitting parameter.

Furthermore, the probability density function $P_g(s_{ab})$ for the normalized speckle intensity is given by
\begin{equation}
    P_g(s_{ab}) = \int_{0}^{\infty} ds_a P_g(s_a)  \frac{e^{-s_{ab}/s_a}}{s_a}.
\end{equation}
The latter expression reveals that the speckle intensity distribution differs from the Rayleigh law, $P(s_{ab})=e^{-s_{ab}}$, only if the normalized total transmission $s_a$ significantly deviates from its mean $\left< s_a \right>=1$.  

\section{Transport mean free path}\label{sec:TMFP}
\begin{figure}
    \centering
    \includegraphics{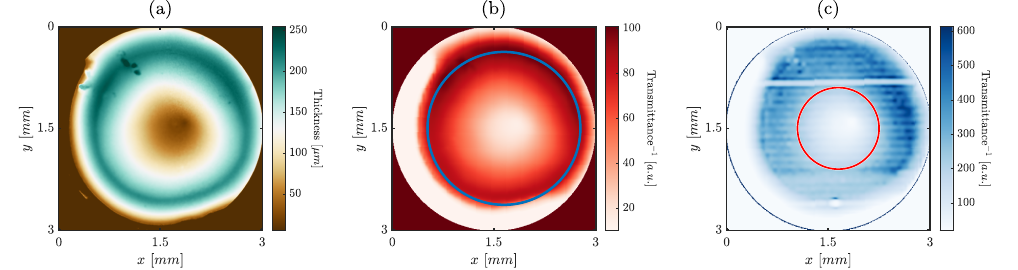}
    \caption{Thickness and inverse transmittance maps of the sample used for the characterization of the transport mean free path. (a) Thickness map measured with a profilometer. (b, c) Inverse transmittance measured at (b) $950\,\mathrm{nm}$ and (c) $450\,\mathrm{nm}$. The circles represent the area used for the analysis of $\ell_t$.
    }
    \label{fig:TMFP_maps_figure}
\end{figure}

We characterized the transport mean free path $\ell_t$ by measuring the scaling of the total transmission with the sample thickness.
The relation between thickness and transport mean free path in absence of absorption follows~\cite{reufer2007transport, schertel2019structural}
\begin{equation}
    T = \frac{ \ell_t + z_0  }{ L + 2z_0 }
    \label{eq:transmittance}
\end{equation}
where $T$ is the transmittance, $L$ is the local thickness of the sample, and $z_0$ the extrapolation length.
The expression of $z_0$ is
\begin{equation}
    z_0 = \frac{2}{3} \left( \frac{1+R_i}{1-R_i} \right) \ell_t = \beta \ell_t\;.
    \label{eq:z0_expression}
\end{equation}
with $R_i$ the internal reflectivity.
The internal reflectivity can be derived from the effective refractive index of the slab $n_{\mathrm{eff}}$. In practice, we use the following approximate expression which is a polynomial fit of the analytical expression~\cite{contini1997photon}
\begin{equation}
\begin{aligned}
    \frac{1+R_i}{1-R_i} &= 504.332889 - 2641.0021n_\mathrm{eff} + 5923.699064n_\mathrm{eff}^2 - 7376.355814n_\mathrm{eff}^3 + \\
 & \qquad + 5507.53041n_\mathrm{eff}^4 - 2463.357945n_\mathrm{eff}^5 + 610.965647n_\mathrm{eff}^6 - 64.8047n_\mathrm{eff}^7 \;.
\end{aligned}
\label{eq:A}
\end{equation}
We obtain the effective refractive index from the Maxwell-Garnett mixing rule~\cite{Markel:16}
\begin{equation}
  n_{\text{eff}}^2 = \frac{1 + \frac{1 + 2\mathrm{ff}}{3}(n^2_{\text{LNO}} - 1)}{1 + \frac{1 - \mathrm{ff}}{3}(n^2_{\text{LNO}} - 1)},
  \label{eq:neff}
\end{equation}
where $n_\mathrm{LNO}$ is the average refractive index of bulk lithium niobate, and $\mathrm{ff}$ is the filling fraction of the nanocubes in the composite.
We calculate the filling fraction by comparing the mass $m$ of the sample, and its volume $V$, estimated from the profilometer map (Fig.~\ref{fig:TMFP_maps_figure}a), using the formula
\begin{equation}
    \mathrm{ff} = \frac{m}{V \rho_\mathrm{LNO}} \approx 0.55\;,
    \label{eq:filling_fraction}
\end{equation}
with $\rho_\mathrm{LNO}$ the density of lithium niobate.
Combining Eqs.~(\ref{eq:z0_expression}-\ref{eq:filling_fraction}) we obtain the value of $\beta$.
Given $\beta$, we recast Eq.~(\ref{eq:transmittance}) to highlight the linear relation between the inverse transmission and the transport mean free path
\begin{equation}
    \frac{1}{T} = \frac{1}{\ell_t + z_0} L + \frac{2z_0}{\ell_t + z_0}
    = \frac{1}{\ell_t (1 + \beta)} L + \frac{2\beta}{1+\beta}\;,
    \label{eq:photonic_ohm_law}
\end{equation}
which has $\ell_t$ as the only free parameter. 
Therefore, we can derive $\ell_t$ from the slope of the relation between the local thickness of the material $L$ and the inverse transmittance.

\begin{figure}
    \centering
    \includegraphics{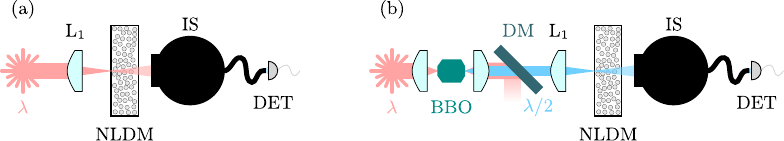}
    \caption{Experimental setup for the characterization of the transmittance. (a) For the measurement at $950\,\mathrm{nm}$, the laser light is directly focused on the disordered medium. (b) Light at $450\,\mathrm{nm}$ is obtained by upconverting the laser with a BBO crystal.  Then, the fundamental light is filtered with a dichroic mirror, and the SH is focused on the nonlinear disordered sample. Finally, the scattered light is collected in both cases by an integrating sphere, and directed towards a detector.
    L: lens; DM: dichroic mirror; NLDM: nonlinear disordered medium; IS: integrating sphere; DET: detector.}
    \label{fig:TMFP_setup}
\end{figure}

\begin{figure}
    \centering
    \includegraphics{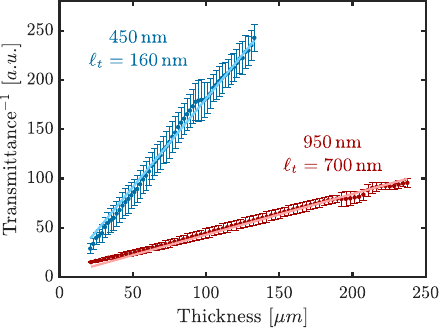}
    \caption{Measured relation between thickness and inverse transmittance at $450 \, \mathrm{nm}$ and $950 \, \mathrm{nm}$. The dots and the errorbars represent the mean values and the standard deviation, respectively. From the slope of the linear fit (solid lines), we extract the transport mean free paths $\ell_t$ according to Eq.~(\ref{eq:photonic_ohm_law}).
    }
    \label{fig:TMFP_plots}
\end{figure}

To this aim, we experimentally measured the thickness map by profilometry (see Fig.~\ref{fig:TMFP_maps_figure}a), and the transmittance of the disordered sample with the experimental setup sketched in Fig.~\ref{fig:TMFP_setup}.
For the measurement of the transmittance at the fundamental wavelength ($\lambda_\mathrm{F} = 950\,\mathrm{nm}$), the laser light is directly focused on the nonlinear disordered medium (Fig.~\ref{fig:TMFP_setup}a).
For the measurement at the SH wavelength ($\lambda_\mathrm{SH} = 450\,\mathrm{nm}$), we first needed to frequency convert the fundamental beam with a BBO crystal.
We then filtered the fundamental light with a dichroic mirror, and we focused the SH beam on the sample (Fig.~\ref{fig:TMFP_setup}b).
In both cases, the scattered light is collected with an integrating sphere, and then guided with a multimode fiber to a powermeter.
We subtracted the background intensity (due to stray light reaching the integrating sphere), and we normalized the measured transmittance values with the intensity transmitted through the bare glass coverslip.
By moving the sample in the plane orthogonal to the illuminating beam, we obtained the 2D map of the sample inverse transmittance at the fundamental and SH wavelength (reported in Figs.~\ref{fig:TMFP_maps_figure}b and c, respectively).
By matching the profilometer and the inverse transmittance map, we obtain the relation between the thickness of the sample and the inverse transmittance (Fig.~\ref{fig:TMFP_plots}).
The areas evaluated for the fit are highlighted with circles in Figs.~\ref{fig:TMFP_maps_figure}b, c.
Note that, due to the lower power available at the SH wavelength, the transmittance measurements at high thicknesses are strongly affected by noise, thus we did not include them in the analysis of the transport mean free path.

We fit the measured data with Eq.~(\ref{eq:photonic_ohm_law}), resulting in the following values of transport mean free paths:
\begin{itemize}
    \item $\ell_t = 700\,\mathrm{nm}$ at $\lambda = 950\,\mathrm{nm}$,
    \item $\ell_t = 160\,\mathrm{nm}$ at $\lambda = 450\,\mathrm{nm}$.
\end{itemize}

The transport mean free path is relevant for the characterization of the mesoscopic effects, because it limits the minimum measurable conductance.
In fact, the light incident on the slab does not contribute to the long-range correlations before it is scattered at least once, linking the intensity fluctuations to the scattering mean free path~\cite{Scheffold1997}.
Combining Eq.~(\ref{eq:variance_and_g}), valid for Gaussian illumination, with the heuristic expression of the intensity fluctuation $C_2=\left<s_a^2\right>-1$ established in Ref.~\cite{Scheffold2002} for small beam diameter, we obtain the following expression of the conductance $g$:
\begin{equation}
    g = \frac{1}{3} \frac{1}{\langle s_a^2 \rangle -1} \approx \frac{2}{3} \left( \frac{8}{9} \right)^2  \frac{\left[ 1 + (9/32) \, w/\ell_t   \right]^2}{1+ (3/16) \, w/l_t } (k \ell_t )^2,
    \label{eq:g_and_lt}
\end{equation}
where $w$ is the input beam width.
For $w \to 0$ we then obtain the minimum measurable conductance $g_\mathrm{min}$, which in our case is
\begin{itemize}
    \item $g_\mathrm{min}(950 \, \mathrm{nm} ) \approx 11.3$,
    \item $g_\mathrm{min}(450 \, \mathrm{nm} ) \approx 2.6$.
\end{itemize}
It is noteworthy that, to obtain a conductance value of $g=6.5$ in Eq.~(\ref{eq:g_and_lt}), we should consider an input width of $w = 600\,\mathrm{nm}$, which corresponds well to the transverse width of the source intensity profile reported in the main text (Fig. 4a on the left).

\section{Transport models in the diffusive regime}\label{sec:transport_model}

To compare the linear propagation of light at frequency $2\omega$ with the propagation of SH light generated within the disordered medium at the same frequency, we compute the mean intensity for both cases, using radiative transport models. The model for SH light, detailed in Ref.~\cite{samanta_speckle_2022}, has been shown to quantitatively match simulations of the microscopic wave equation that incorporate the SH generation process.
 Here we solve both propagation models in the diffusive limit where the transport mean free path is much smaller than the medium thickness, using focused beams for illumination.   

 We consider a disordered slab translation-invariant in $x$ and $y$ directions and of thickness $L$ in direction $z$. Scattering is supposed to be isotropic such that the scattering and transport mean-free paths are equals (i.e., $\ell_s=\ell_t$ and anisotropy factor $g=0$) and there is no absorption. When this slab is illuminated by a Gaussian beam from $z<0$ at a frequency $\omega$, the mean ballistic intensity at depth $z\ge0$ and frequency $\omega$ takes the following expression:
\begin{equation}
    I_b(\mathbf{R},z,\omega)=I_0\exp\left[-\frac{2R^2}{w^2}-\frac{z}{\ell_s(\omega)}\right],
\end{equation}
where $I_0$ is the intensity of the incident beam of waist $w$, and $\mathbf{R} =(x,y)$ is the transverse coordinate, with $R=\sqrt{x^2+y^2}$. This ballistic intensity is a source term for the diffuse intensity $I_d$ which obeys the stationary diffusion equation
\begin{equation}
    -\frac{\ell_s(\omega)^2}{3}\nabla^2 I_d(\mathbf{R},z,\omega)=I_b(\mathbf{R},z,\omega)
\end{equation}
with the boundary conditions
\begin{align}
    I_d(\mathbf{R},z=0,\omega)-z_0(\omega)\frac{\partial I_d}{\partial z}(\mathbf{R},z=0,\omega) & =0,
\\
    I_d(\mathbf{R},z=L,\omega)+z_0(\omega)\frac{\partial I_d}{\partial z}(\mathbf{R},z=L,\omega) & =0,
\end{align}
where $z_0(\omega)=2\ell_s(\omega)/3$ is the extrapolation length. This set of equations can be easily solved applying a Fourier transform with respect to $\mathbf{R}$. In particular, the Green function, solution of
\begin{equation}
    -\left(\frac{\partial^2}{\partial z^2}-q^2\right) G_d(\mathbf{q},z,z',\omega)=\delta(z-z')
\end{equation}
with
\begin{align}
    G_d(\mathbf{q},z=0,z',\omega)-z_0(\omega)\frac{\partial G_d}{\partial z}(\mathbf{q},z=0,z',\omega) & =0,
\\
    G_d(\mathbf{q},z=L,z',\omega)+z_0(\omega)\frac{\partial G_d}{\partial z}(\mathbf{q},z=L,z',\omega) & =0,
\end{align}
is given by
\begin{equation}
\label{EqGreenDiff}
    G_d(\mathbf{q},z,z',\omega)=
    \begin{cases}\displaystyle
        \frac{1}{q}
            \frac{\left[ 
                \sinh(qz)+qz_0(\omega)\cosh(qz)
            \right]\left[
                \sinh\{q(L-z')\}+qz_0(\omega)\cosh\{q(L-z')\}
            \right]}
            {\left[1+q^2z_0(\omega)^2\right]\sinh(qL)+2qz_0(\omega)\cosh(qL)}
            & \text{for $z<z'$},
    \\\displaystyle
        \frac{1}{q}
            \frac{\left[ 
                \sinh(qz')+qz_0(\omega)\cosh(qz')
            \right]\left[
                \sinh\{q(L-z)\}+qz_0(\omega)\cosh\{q(L-z)\}
            \right]}
            {\left[1+q^2z_0(\omega)^2\right]\sinh(qL)+2qz_0(\omega)\cosh(qL)}
            & \text{for $z>z'$}.
    \end{cases}
\end{equation}
From this expression, the diffuse intensity is given by
\begin{equation}
    I_d(\mathbf{q},z,\omega)=\frac{3}{\ell_s(\omega)^2}\int_0^L \ud z' G_d(\mathbf{q},z,z',\omega)I_b(q,z',\omega),
\end{equation}
where $I_b(\mathbf{q},z',\omega)$ is the Fourier transform of $I_b(\mathbf{R},z',\omega)$. We can write it explicitly as
\begin{equation}
I_d(\mathbf{q},z,\omega)=I_0\frac{3\pi w^2}{2\ell_s(\omega)^2}e^{-q^2w^2/8}\bar{G}_d(\mathbf{q},z,\omega),
\end{equation}
where the integrated Green's function, $\bar{G}_d(\mathbf{q},z,\omega)=\int_0^L \ud z' G_d(\mathbf{q},z,z',\omega)e^{-z'/\ell_s(\omega)}$, is given by 
\begin{align}
\bar{G}_d(\mathbf{q},z,\omega)= & \frac{l_s(\omega)^2}{q^2l_s(\omega)^2-1}
\left[
e^{-z/l_s(\omega)} 
\right.
\nonumber
\\
&
\left.
- \frac{
e^{-L/l_s(\omega)}[2ql_s(\omega)\cosh(qz)+3\sinh(qz)] + 10ql_s(\omega)\cosh\{q(L-z)\}+ 15\sinh\{q(L-z)\} 
}
{12 q l_s(\omega)\cosh(qL)+[9+4q^2l_s(\omega)^2]\sinh(qL)}
\right].
\end{align}
Taking the inverse Fourier transform, we  obtain
\begin{equation}
I_d(\mathbf{R},z,\omega)=I_0\frac{3 w^3}{4\ell_s(\omega)^2}\int_0^\infty \ud q e^{-q^2w^2/8}\bar{G}_d(\mathbf{q},z,\omega)q J_0(qR),
\end{equation}
where $J_0$ is the Bessel function of the first kind of order zero. Finally, the total mean intensity for linear propagation at frequency $\omega$ reads 
\begin{equation}
I_\text{lin}(\mathbf{R},z,\omega)=I_b(\mathbf{R},z,\omega)+I_d(\mathbf{R},z,\omega).    
\end{equation}
The source term represented in the left part of Fig.~4c of the main text corresponds to $I_b(\mathbf{R},z,2\omega)$, while the total intensity profile shown in the right part is $I_\text{lin}(\mathbf{R},z,2\omega)$, evaluated at $x=0$. Parameters of the calculation are chosen close to the experimental values: the waist $w\simeq 0.41\,\mu$m of the incident beam is such that its FWHM, $\sqrt{2\ln(2)}w$, is equal to $(\lambda_F/2)/2\text{NA}$ with $\lambda_F=0.976\,\mu$m and $\text{NA}=0.5$; the mean free path is $\ell_s(2\omega)=0.2\,\mu$m; the sample thickness is $L=10\,\mu$m.   

Regarding SH light propagating at frequency $2\omega$, it is possible to show that the diffuse mean intensity still verifies a diffusion equation, with a source term that involves the square of the mean intensity of light propagating at frequency $\omega$. For isotropic scattering, we have~\cite{samanta_speckle_2022}
\begin{equation}
    -\frac{\ell_s(2\omega)}{3}\nabla^2 I_d(\mathbf{R},z,2\omega)=\mathcal{M}_0(\mathbf{R},z,\omega),
\end{equation}
with
\begin{equation}
\label{EqSourceSHG}
    \mathcal{M}_0(\mathbf{R},z,\omega)=\frac{\alpha}{4\pi}I_\text{lin}(\mathbf{R},z,\omega)^2.
\end{equation}
Here $\alpha$ is a parameter proportional in particular to the SH susceptibility $\chi^{(2)}$, whose explicit value does not affect the profile of the SH light propagation. With the same boundary conditions as in the fundamental case, the solution is given by
\begin{equation}
\label{EqDiffSHG}
    I_d(\mathbf{R},z,2\omega)=\frac{3}{\ell_s(2\omega)}\int_0^L\ud z'\iint \ud \mathbf{R}'
    \,
    G_d(\mathbf{R}-\mathbf{R}',z,z',2\omega)\mathcal{M}_0(\mathbf{R}',z',\omega),
\end{equation}
where $G_d(\mathbf{R},z,z',\omega)$ is the inverse Fourier transform of the expression given in Eq.~\eqref{EqGreenDiff}. The source term shown in the left part of Fig.~4a of the main text corresponds to Eq.~\eqref{EqSourceSHG} evaluated at $x=0$, with $w\simeq 0.83\,\mu$m and $\ell_s(\omega)=0.7\,\mu$m. On the other hand, the right part of Fig.~4a shows the diffusive SH intensity given by Eq.~\eqref{EqDiffSHG}, computed with the values of the parameters mentioned above. 

\bibliography{references}